\documentclass{jpp}
\usepackage{graphicx}

\usepackage[utf8]{inputenc}
\usepackage[T1]{fontenc}
\usepackage{amsmath}

\usepackage{xcolor}
\usepackage[normalem]{ulem}
\usepackage{cancel}

\newcommand{\del}[1]{}

\shorttitle{Axisymmetric adjoint reduction}
\shortauthor{F. Tanji et al.}

\title{Axisymmetric reduction of the adjoint operator in ideal MHD equilibrium}

\author{Fumiya Tanji\aff{1}
  \corresp{\email{tanji.fumiya.56e@st.kyoto-u.ac.jp}}
  , Akinobu Matsuyama\aff{1}
  , Akio Sanpei\aff{2}
  , Sadao Masamune\aff{2,3}
  , Yasuaki Kuroe\aff{4}
  \and Yuji Nakamura\aff{1}}

\affiliation{\aff{1}Graduate School of Energy Science, Kyoto University, Kyoto 611-0011, Japan
             \aff{2}Department of Electronics, Kyoto Institute of Technology, Kyoto 606-8585, Japan
             \aff{3}College of Engineering, Chubu University, Aichi 487-8502, Japan
             \aff{4}Mobility Research Center, Doshisha University, Kyotanabe, Kyoto 610-0384, Japan}

\begin{document}

\maketitle

\begin{abstract}
Adjoint formulations of magnetohydrodynamic (MHD) equilibrium are presented for sensitivity analysis and optimization-based equilibrium calculations. 
We derive the adjoint of the three-dimensional equilibrium equations under fixed-boundary conditions and analyse its axisymmetric reduction. 
The reduced operator coincides with the perturbed Grad–Shafranov operator, thereby making explicit the connection between the three-dimensional adjoint construction and the
axisymmetric adjoint problem.
The resulting framework provides a basis for relating adjoint structures in both
axisymmetric and fully three-dimensional ideal MHD equilibrium.
\end{abstract}

\section{Introduction}

Ideal magnetohydrodynamic (MHD) equilibrium provides a fundamental description of magnetically confined plasmas \citep{Freidberg2014}. 
Equilibrium calculations for finite-$\beta$ plasmas are essential in the design of experimental devices, and equilibrium reconstruction from diagnostic data is indispensable for understanding the internal state of the plasma \citep{Lao1985, Blum2012}, such as the current and plasma profiles. In axisymmetric configurations such as tokamaks, MHD equilibrium is obtained by solving the Grad-Shafranov equation \citep{Grad1958,Shafranov1958}, whereas in three-dimensional system it is generally computed using numerical equilibrium solvers,  such as VMEC \citep{Hirshman1983} and HINT \citep{HINT1,HINT2}.

Equilibrium calculations for the experiment design and equilibrium reconstruction can be formulated as constrained optimization problems that extremise a prescribed objective function. With advances in computational capabilities, the optimization-based equilibrium calculations have recently been applied to both axisymmetric and three-dimensional configurations \citep{Sanpei2021,Sanpei2023,Landreman2022}. In such calculations, performing sensitivity analysis of equilibrium solutions with respect to a large number of parameters can be computationally expensive. By exploiting the mathematical structure of the ideal MHD equilibrium problem, however, the adjoint problem can be formulated to evaluate these sensitivities efficiently. For axisymmetric systems, the adjoint method has been employed for the reconstruction of Grad-Shafranov equilibria in tokamaks and RFPs \citep{Sanpei2021,Sanpei2023}. The adjoint formulation has further been used to evaluate shape sensitivities in three-dimensional stellarator optimisation, based on the linearisation of the equilibrium operator \citep{Paul2018,Paul2020,Antonsen2019,Gaur2023}. These parallel developments lead to the question of how the adjoint structure of the full three-dimensional equilibrium problem reduces, under the assumption of axisymmetry, to adjoint formulations derived from the Grad-Shafranov equation.

The Grad-Shafranov equation represents the axisymmetric reduction of the ideal MHD equilibrium equations. It is therefore natural to expect that the adjoint formulation of the full three-dimensional problem should likewise reduce, under the assumption of axisymmetry, to the adjoint of the Grad-Shafranov equation. To the best of our knowledge, however, the precise relationship between the adjoint formulation of the full three-dimensional problem and that for the Grad-Shafranov equation has not been explicitly demonstrated in previous studies.

The purpose of the present work is to clarify this relationship. Starting from the three-dimensional ideal MHD equilibrium equations, we derive the corresponding adjoint operator without assuming symmetry.
We then impose axisymmetry and perform an explicit reduction of the three-dimentional adjoint system. We show that the resulting adjoint operator coincides with the perturbed Grad-Shafranov operator \citep{Zakharov1999}, which governs infinitesimal variations of the equilibrium flux function.

Understanding this result requires attention to the operator relations defined with respect to the inner product weighted by the geometric factor $1/R^2$. This factor arises in the variational formulation of axisymmetric magnetic energy and defines the appropriate inner-product structure of the Grad-Shafranov operator \citep{Lao1981,Freidberg2015}. Although this connection may appear formally straightforward, the direct correspondence between the three-dimensional adjoint variables and their axisymmetric counterparts has not, to our knowledge, been made explicit in the literature.

In \cite{Sanpei2021,Sanpei2023}, an adjoint formulation of the Grad-Shafranov equation was derived without retaining the geometric weight $1/R^2$. Adopting such an inner-product structure does not pose any practical difficulties within the framework of constrained optimization for numerical solutions of partial differential equations. However, the self-adjointness of the linear MHD operator is then lost, and the structural relation between the full three-dimensional problem and its axisymmetric reduction becomes obscured. By contrast,
the present formulation makes explicit an operator-level correspondence between the homogeneous adjoint system of the full three-dimensional equilibrium problem and its axisymmetric reduction. We show that the axisymmetric reduction of the three-dimensional homogeneous adjoint system recovers the same reduced scalar operator as the adjoint formulation based on the Grad-Shafranov equation. This correspondence clarifies how the axisymmetric adjoint structure is represented within the three-dimensional adjoint formulation, and suggesting a useful reference framework for sensitivity analysis of weakly non-axisymmetric or quasi-axisymmetric configurations.

The paper is organised as follows. 
In \S\ref{sec:Adjoint formulation for constrained equilibrium problems} we describe the adjoint formulation for constrained equilibrium problems. 
In \S\ref{sec:Adjoint equations for three-dimensional ideal MHD equilibrium} we derive the adjoint equations for three-dimensional ideal MHD equilibria under fixed-boundary conditions. 
In \S\ref{sec:Axisymmetric reduction of the three-dimensional adjoint operator} we perform the axisymmetric reduction and demonstrate the equivalence with the perturbed Grad–Shafranov operator. 
In \S\ref{sec:Discussion}, we discuss the implications of this reduction for interpreting weakly non-axisymmetric equilibria within the three-dimensional adjoint framework.
The conclusions are given in \S\ref{sec:Conclusion}.

\section{Adjoint formulation for constrained equilibrium problems}\label{sec:Adjoint formulation for constrained equilibrium problems}

The MHD equilibrium equations constitute a nonlinear boundary-value problem depending on external parameters. To clarify the operator structure underlying the associated adjoint formulation, we introduce a general framework for constrained equilibrium problems.

Let $\rho$ denote external parameters and $y$ denote the state variables describing the equilibrium. 
We assume that the equilibrium satisfies the operator equation
\begin{align*}
    f(y,\rho)=0 ,
\end{align*}
where $f$ represents a (generally nonlinear) differential operator. 
We consider a functional
\begin{align*}
    j = j(y,\rho),
\end{align*}
which depends on both the equilibrium solution and external parameters.

We introduce the Lagrangian
\begin{align}
    \mathcal{L}(y,\lambda;\rho)
    =
    j(y,\rho)
    +
    ( \lambda, f(y,\rho) ) .
\end{align}
where $\lambda$ denotes the adjoint variable (Lagrange multiplier)
associated with the constraint $f(y,\rho)=0$, and $(\cdot,\cdot)$denotes the inner product.
Requiring stationarity with respect to $\lambda$ and $y$ yields
\begin{equation*}
    f(y,\rho)=0 ,
\end{equation*}
and
\begin{equation}
    \frac{\partial j}{\partial y}
    +
    \Big( \frac{\partial f}{\partial y} \Big)^* \lambda
    = 0, \label{eq:general adjoint}
\end{equation}
where $(\partial f/\partial y)^*$ denotes the adjoint of the linearised operator $(\partial f/\partial y)$ with respect to a given inner product.

Linearising the constraint gives
\begin{align*}
    \frac{\partial f}{\partial y}\,\delta y
    +
    \frac{\partial f}{\partial \rho}\,\delta \rho
    = 0 ,
\end{align*}
while the variation of $j$ is given by
\begin{align*}
    \delta j
    =
    \frac{\partial j}{\partial y}\,\delta y
    +
    \frac{\partial j}{\partial \rho}\,\delta \rho .
\end{align*}
Eliminating $\delta y$ using the adjoint equation \eqref{eq:general adjoint} yields
\begin{align}
    \delta j
    =
    \Big[
        \frac{\partial j}{\partial \rho}
        +
        \Big( \frac{\partial f}{\partial \rho} \Big)^* \lambda
    \Big]
    \delta \rho .
\end{align}
Thus, the adjoint variable encodes the action of the adjoint of the linearised equilibrium operator without requiring an explicit solution of the linearised state equation.

\section{Adjoint equations for three-dimensional ideal MHD equilibrium}\label{sec:Adjoint equations for three-dimensional ideal MHD equilibrium}

We now apply the general formulation of \S\ref{sec:Adjoint formulation for constrained equilibrium problems} to the three-dimensional ideal MHD equilibrium. 
We consider a static equilibrium in a bounded domain 
$\Omega \subset \mathbb{R}^3$ with fixed boundary 
$\Gamma = \partial \Omega$.

The state variables are the magnetic field $\boldsymbol{B}$, 
current density $\boldsymbol{J}$, and pressure $p$. 
The governing equations are
\begin{subequations}
\begin{align}
    \nabla p &= \boldsymbol{J} \times \boldsymbol{B}, \\
    \mu_0 \boldsymbol{J} &= \nabla \times \boldsymbol{B}, \\
    \nabla \cdot \boldsymbol{B} &= 0 .
\end{align}
\end{subequations}

On the boundary $\Gamma$, we impose the perfectly conducting wall condition
\begin{align*}
    \boldsymbol{B} \cdot \boldsymbol{n} = 0 .
\end{align*}

In the notation of \S\ref{sec:Adjoint formulation for constrained equilibrium problems}, the state vector is
\[
    y = (\boldsymbol{J}, \boldsymbol{B}, p),
\]
and the constraint $f(y)=0$ corresponds to the above equilibrium equations.
The corresponding linearised variation is $\delta y = (\delta\boldsymbol{J},~\delta\boldsymbol{B},~\delta p)$, which satisfies the linearised equilibrium constraints
\begin{align*}
    \nabla\delta p &= \delta\boldsymbol{J}\times\boldsymbol{B} + \boldsymbol{J}\times\delta\boldsymbol{B} \,\\
    \mu_0\delta\boldsymbol{J} &= \nabla\times\delta\boldsymbol{B} \,\\
    \nabla\cdot\delta\boldsymbol{B} &= 0 .
\end{align*}
This point should be distinguished from the usual Lagrangian-displacement formulation of ideal MHD. In the energy principle \citep{Bernstein1958}, variations are generated by a plasma displacement $\boldsymbol{\xi}$, and the perturbations of $p$ and $\boldsymbol{B}$ are restricted by the ideal-MHD constraints. By contrast, the present formulation treats $(\boldsymbol{J},~\boldsymbol{B},~p)$ as independent state variables subject to the static equilibrium equations above. This distinction is relevant in sensitivity analysis and equilibrium reconstruction, where variations of the equilibrium fields need not be restricted to ideal-MHD displacement perturbations. The adjoint variables introduced below are therefore the Lagrange multipliers dual to the static equilibrium constraints, rather than physical displacement variables.

To derive the adjoint operator associated with the linearised equilibrium system, we introduce the Lagrangian
\begin{align}
    \mathcal{L}
    &= \int_\Omega \mathcal{F}\, dV
    + \int_\Omega 
      \boldsymbol{\alpha} \cdot 
      (\nabla p - \boldsymbol{J} \times \boldsymbol{B}) \, dV 
    \nonumber \\
    &\quad
    + \int_\Omega 
      \boldsymbol{\beta} \cdot 
      (\mu_0 \boldsymbol{J} - \nabla \times \boldsymbol{B}) \, dV 
    + \int_\Omega 
      \gamma \, (\nabla \cdot \boldsymbol{B}) \, dV ,
\end{align}
where $\mathcal{F}$ denotes the integrand of the objective functional $j$, and $\boldsymbol{\alpha}$, $\boldsymbol{\beta}$ and $\gamma$ are adjoint variables.

Taking variations with respect to 
$\boldsymbol{J}$, $\boldsymbol{B}$ and $p$, 
and applying standard vector identities together with the divergence theorem, 
we obtain
\begin{align*}
    \delta \mathcal{L}
    &= \int_\Omega 
       \delta \boldsymbol{J} \cdot
       \Big[
           \frac{\delta \mathcal{F}}{\delta \boldsymbol{J}}
           - \boldsymbol{B} \times \boldsymbol{\alpha}
           + \mu_0 \boldsymbol{\beta}
       \Big] dV
       \nonumber \\
    &+ \int_\Omega 
       \delta \boldsymbol{B} \cdot
       \Big[
           \frac{\delta \mathcal{F}}{\delta \boldsymbol{B}}
           - \boldsymbol{\alpha} \times \boldsymbol{J}
           - \nabla \times \boldsymbol{\beta}
           - \nabla \gamma
       \Big] dV
       \nonumber \\
    &+ \int_\Omega 
       \delta p
       \Big[
           \frac{\delta \mathcal{F}}{\delta p}
           - \nabla \cdot \boldsymbol{\alpha}
       \Big] dV
       \nonumber \\
    &+ \oint_\Gamma
       \Big[
           \delta p\, (\boldsymbol{\alpha} \cdot \boldsymbol{n})
           - \delta \boldsymbol{B} \cdot
             (\boldsymbol{\beta} \times \boldsymbol{n})
           + \gamma \, (\delta \boldsymbol{B} \cdot \boldsymbol{n})
       \Big] dS .
\end{align*}

We assume a fixed-boundary setting in which
\begin{align*}
    \delta p = 0,
    \qquad
    \delta \boldsymbol{B} \cdot \boldsymbol{n} = 0
    \qquad \text{on } \Gamma .
\end{align*}
Under these assumptions, the remaining boundary contribution is removed by imposing the adjoint boundary condition
\begin{align*}
    \boldsymbol{\beta} \times \boldsymbol{n} = 0
    \qquad \text{on } \Gamma .
\end{align*}

Requiring the volume terms to vanish for arbitrary variations yields the adjoint equations
\begin{subequations}
\begin{align}
    \frac{\delta \mathcal{F}}{\delta \boldsymbol{J}}
    - \boldsymbol{B} \times \boldsymbol{\alpha}
    + \mu_0 \boldsymbol{\beta}
    &= 0, \\
    \frac{\delta \mathcal{F}}{\delta \boldsymbol{B}}
    - \boldsymbol{\alpha} \times \boldsymbol{J}
    - \nabla \times \boldsymbol{\beta}
    - \nabla \gamma
    &= 0, \\
    \frac{\delta \mathcal{F}}{\delta p}
    - \nabla \cdot \boldsymbol{\alpha}
    &= 0 .
\end{align}
\end{subequations}
The scalar multiplier $\gamma$ appears only through its gradient and is defined up to an additive constant.
It should be noted that no magnetic-surface coordinates, flux functions, or symmetry assumptions have been introduced up to this point. Thus, the adjoint equations above correspond to the three-dimensional static MHD equilibrium constraints written in the state variables $(\boldsymbol{J},~\boldsymbol{B},~p)$. In this formulation, $\boldsymbol{J}$ has been treated as an independent variable to obtain a first-order constrained formulation. It remains constrained by Amp\`ere's law.

To isolate the intrinsic operator structure independently of any particular functional, we now consider the homogeneous case $\mathcal{F}=0$. 
The adjoint system then reduces to
\begin{subequations}
\label{eq:adjoint_system}
\begin{align}
    \boldsymbol{B} \times \boldsymbol{\alpha}
    - \mu_0 \boldsymbol{\beta} &= 0, \label{eq:adjoint1} \\
    \boldsymbol{\alpha} \times \boldsymbol{J}
    + \nabla \times \boldsymbol{\beta}
    + \nabla \gamma &= 0, \label{eq:adjoint2} \\
    \nabla \cdot \boldsymbol{\alpha} &= 0 . \label{eq:adjoint3}
\end{align}
\end{subequations}

In notation of \S\ref{sec:Adjoint formulation for constrained equilibrium problems}, the equilibrium equations are written as $f(y)=0$, and the linearised operator at an equilibrium $y_0$ is given by $L = Df[y_0]$. The adjoint system takes the compact form
\begin{align*}
    L^* \Lambda = - \frac{\delta \mathcal{F}}{\delta y},
\end{align*}
with $\Lambda = (\boldsymbol{\alpha}, \boldsymbol{\beta}, \gamma)$.
In particular, the homogeneous adjoint system corresponds to
\begin{align}
L^* \Lambda = 0 ,
\end{align}
which characterises the kernel of the adjoint of the linearised equilibrium operator.

The homogeneous adjoint system has a structural analogy with frozen-in perturbations. The vector field $\boldsymbol{\alpha}$ plays an incompressible displacement-like role, $-\mu_0\boldsymbol{\beta}$ has a vector-potential-perturbation-like character, and $\gamma$ appears as a gauge-like scalar in the magnetic representation. Under this analogy, the adjoint equations give
\begin{align*}
    &\delta\boldsymbol{B}_{\rm ad} = \nabla\times(\boldsymbol{\alpha}\times\boldsymbol{B}), \\
    &\delta\boldsymbol{J}_{\rm ad} = \nabla\times(\boldsymbol{\alpha}\times\boldsymbol{J}),
\end{align*}
where the adjoint-induced fields have been defined as $\delta\boldsymbol{B}_{\rm ad} \equiv \nabla \times (-\mu_0\boldsymbol{\beta}),~\delta\boldsymbol{J}_{\rm ad} \equiv \frac{1}{\mu_0}\nabla\times\delta\boldsymbol{B}_{\rm ad}$. In this sense, both the magnetic field and the current field admit frozen-in-like representations in the homogeneous adjoint system.

\section{Axisymmetric reduction of the three-dimensional adjoint operator}\label{sec:Axisymmetric reduction of the three-dimensional adjoint operator}

We now consider the homogeneous adjoint system under the assumption that the equilibrium and adjoint fields are axisymmetric.
Let $\phi$ denote the toroidal angle. 
Under axisymmetry, divergence-free vector fields may be written in the standard representation \citep{Dhaeseleer1991}
\begin{subequations}
\label{axisymmetric representation}
\begin{align}
\boldsymbol{B}
&= B_\phi \nabla\phi
   + \nabla\phi\times\nabla\psi , \\
\boldsymbol{\alpha}
&= \alpha_\phi \nabla\phi
   + \nabla\phi\times\nabla\alpha_1 ,
\end{align}
\end{subequations}
where all scalar functions are independent of $\phi$. 
Here $\psi$ denotes the poloidal flux function, so that magnetic surfaces are given by $\psi=\mathrm{const}$.

Using Amp\' ere's law,
\begin{align}
\mu_0 \boldsymbol{J}
= \nabla\times\boldsymbol{B}
= \nabla B_\phi \times \nabla\phi
  + \triangle^* \psi \, \nabla\phi ,
\end{align}
where
\begin{align*}
\triangle^* \psi
= R^2 \nabla\cdot\!\left(\frac{1}{R^2}\nabla\psi\right),
\end{align*}
and $|\nabla\phi|^2 = 1/R^2$.

From the first adjoint equation \eqref{eq:adjoint1} we obtain
\begin{align*}
\mu_0 \boldsymbol{\beta}
= \boldsymbol{B}\times\boldsymbol{\alpha}.
\end{align*}
Substituting the representations \eqref{axisymmetric representation} and using vector identities yields
\begin{align}
\mu_0 \boldsymbol{\beta}
=
-\frac{B_\phi}{R^2}\nabla\alpha_1
+ \frac{\alpha_\phi}{R^2}\nabla\psi
+ \eta\,\nabla\phi ,
\end{align}
where
\begin{align}
\eta
= \nabla\alpha_1 \cdot (\nabla\phi\times\nabla\psi).
\end{align}
The scalar quantity $\eta$ admits a simple geometric interpretation.
Since
\begin{align*}
\eta
=
\nabla \alpha_1 \cdot (\nabla\phi \times \nabla\psi)
=
- \nabla\psi \cdot \boldsymbol{\alpha},
\end{align*}
where the toroidal component of $\boldsymbol{\alpha}$ vanishes under axisymmetry, $\eta$ represents the projection of the adjoint field along $\nabla\psi$.
If $\boldsymbol{\alpha}$ is viewed as a displacement-like field, this relation is formally analogous to the perturbation of the poloidal flux in ideal MHD \citep{Paul2020},
\begin{align*}
\delta\psi
=
- \boldsymbol{\xi} \cdot \nabla\psi ,
\end{align*}
so that $\eta$ may be interpreted as a flux-like perturbation associated with axisymmetric displacements.

We now substitute these expressions into the second adjoint equation \eqref{eq:adjoint2} and isolate the scalar structure by projecting along and across $\nabla\phi$.
Taking the cross product of Eq. \eqref{eq:adjoint2} with $\nabla\phi$ eliminates the gradient term $\nabla\gamma$. After straightforward but lengthy algebra, the following expression can be obtained as
\begin{align}
    \frac{\alpha_\phi}{R^2}
    \nabla\phi\times\nabla B_\phi
    + \frac{\triangle^*\psi}{R^2}
    \nabla\phi\times\nabla\alpha_1
    + \frac{1}{R^2}\nabla\eta
    + \nabla\phi\times\nabla\gamma
    = 0 , \label{eq:adjoint_sub_2}
\end{align}
Taking the divergence of Eq. \eqref{eq:adjoint_sub_2} and using axisymmetry yields
\begin{align}
    \nabla\!\left(\frac{\alpha_\phi}{R^2}\right)
    \!\cdot\!
    \nabla\phi\times\nabla B_\phi
    + \nabla\!\left(\frac{\triangle^*\psi}{R^2}\right)
    \!\cdot\!
    \nabla\phi\times\nabla\alpha_1
    + \nabla\cdot\!\left(
        \frac{1}{R^2}\nabla\eta
      \right)
    = 0 . \label{eq:phi_relation_div}
\end{align}

The scalar product of Eq. \eqref{eq:adjoint2} with $\nabla\phi$ yields
\begin{align*}
(\boldsymbol{\alpha}\times\boldsymbol{J})\cdot\nabla\phi
+
(\nabla\times\boldsymbol{\beta})\cdot\nabla\phi
=0,
\end{align*}
since $\nabla\gamma\cdot\nabla\phi=0$ by axisymmetry. Substituting the axisymmetric representations of $\boldsymbol{J}$ and $\boldsymbol{\beta}$ and using 
\(
\nabla\phi\times(\nabla\phi\times\nabla f)
= -|\nabla\phi|^2 \nabla f
\)
with $|\nabla\phi|^2=1/R^2$, one obtains
\begin{align}
R^2 \nabla\!\left(\frac{\alpha_\phi}{R^2}\right)\cdot(\nabla\phi\times\nabla\psi)
+
\frac{2B_\phi}{R}\,
\nabla R\cdot(\nabla\phi\times\nabla\alpha_1)
=0 .
\label{eq:phi_relation}
\end{align}

Using the equilibrium Grad–Shafranov equation
\begin{align*}
\triangle^*\psi
=
-\mu_0 R^2 p'(\psi)
-\frac{1}{2}(F^2)'(\psi),
\qquad
B_\phi=F(\psi),
\end{align*}
Eq. \eqref{eq:phi_relation_div} becomes
\begin{align}
R^2\nabla\!\left(\frac{\alpha_\phi}{R^2}\right)
\!\cdot\!(\nabla\phi\times\nabla B_\phi)
&+
\left(
\triangle^*
+\mu_0 R^2 p''(\psi)
+\frac{1}{2}(F^2)''(\psi)
\right)\eta
\nonumber \\
&+
\frac{(F^2)'(\psi)}{R}\,
\nabla R\cdot(\nabla\phi\times\nabla\alpha_1)
=0 .
\label{eq:div_compact}
\end{align}

Since $B_\phi=F(\psi)$ and $(F^2)'=2F F'$, the first and third terms of Eq. \eqref{eq:div_compact} combine to give $F'$ times \eqref{eq:phi_relation} and therefore vanish. We thus obtain
\begin{align}
\left(
\triangle^*
+\mu_0 R^2 p''(\psi)
+\frac{1}{2}(F^2)''(\psi)
\right)\eta
=0 , \label{eq:perturbed Grad_Shafranov}
\end{align}
which coincides with the perturbed Grad--Shafranov equation \citep{Zakharov1999}.

The above derivation gives an explicit consistency statement for the axisymmetric restriction of the three-dimensional adjoint formulation. Constructing the adjoint before imposing axisymmetry gives the same reduced scalar operator as constructing the adjoint after reduction to the Grad-Shafranov equation. This reduction also clarifies how the reduced scalar variable is represented in the three-dimensional adjoint formulation. Specifically, the scalar adjoint variable is obtained from the three-dimensional adjoint field as
\begin{align*}
    \eta = -\boldsymbol{\alpha}\cdot\nabla\psi ,    
\end{align*}
where $\boldsymbol{\alpha}$ is incompressible in the homogeneous case. Thus, the reduced scalar adjoint variable may be interpreted as the flux-function variation associated with an incompressible axisymmetric adjoint field.

This result also clarifies the structure in the axisymmetric equilibrium and adjoint operators. 
The scalar equation of Eq. \eqref{eq:perturbed Grad_Shafranov} may be written as
\begin{align*}
\mathcal{L}\eta = 0 ,
\end{align*}
where
\begin{align}
\mathcal{L}
=
\triangle^*
+\mu_0 R^2 p''(\psi)
+\frac{1}{2}(F^2)''(\psi),
\qquad
\triangle^*
=
R^2\nabla\cdot\!\left(\frac{1}{R^2}\nabla\right). \label{eq:linearised Grad-Shafranov operator}
\end{align}
To make the adjoint structure explicit, we consider the weighted inner product
\begin{align}
(u,v)_w
:=
\int_\Omega u\,v\,\frac{1}{R^2}\,dV .
\end{align}
For sufficiently smooth scalar functions $u$ and $v$, integration by parts yields
\begin{align*}
(u,\triangle^* v)_w
=
\oint_\Gamma
\frac{u\,\boldsymbol{n}\cdot\nabla v}{R^2}\,dS
-
\int_\Omega
\frac{\nabla u\cdot\nabla v}{R^2}\,dV .
\end{align*}
Exchanging $u$ and $v$ and subtracting gives
\begin{align}
(u,\triangle^* v)_w
-
(\triangle^* u,v)_w
=
\oint_\Gamma
\frac{
u\,\boldsymbol{n}\cdot\nabla v
-
v\,\boldsymbol{n}\cdot\nabla u
}{R^2}\,dS .
\end{align}
Hence $\triangle^*$ is formally self-adjoint with respect to $(\cdot,\cdot)_w$ under boundary conditions that make the boundary term vanish \citep{Lao1981,Freidberg2015}.
Since the remaining terms in $\mathcal{L}$ of Eq. \eqref{eq:linearised Grad-Shafranov operator} act as multiplication operators, the full scalar operator $\mathcal{L}$ is formally self-adjoint in the same weighted space.
Thus, although the present operator is derived from a fully three-dimensional adjoint formulation, its axisymmetric reduction has the same weighted self-adjoint form as the scalar operator governing the marginal stability of the incompressible $n=0$ mode in ideal MHD \citep{Freidberg2015}. 

The weighted inner product used above is not introduced ad hoc, but is motivated by the axisymmetric form of the magnetic energy. Indeed, the poloidal magnetic energy is
$E_{\rm pol}=\int_{\Omega}|\boldsymbol{B}_{\rm pol}|^2/(2\mu_0)\,dV$.
For an axisymmetric magnetic field, the poloidal component can be written as
$\boldsymbol{B}_{\rm pol}=\nabla\psi\times\nabla\phi$, where $\phi$ is the toroidal angle. Since $|\nabla\phi|^2=1/R^2$ and $\nabla\psi\cdot\nabla\phi=0$, this energy becomes
\begin{align*}
    E_{\rm pol}
    =
    \frac{1}{2\mu_0}
    \int_{\Omega}
    \frac{|\nabla\psi|^2}{R^2}\,dV .
\end{align*}
Taking the first variation with respect to $\psi$ and integrating by parts gives, and assuming that the boundary term vanishes, gives
\begin{align*}
    \delta E_{\rm pol}
    =
    -\frac{1}{\mu_0}
    \int_{\Omega}
    \delta\psi\,\triangle^*\psi\,\frac{dV}{R^2}.
\end{align*}
Thus, when the magnetic-energy variation is written in terms of the Grad--Shafranov operator, scalar perturbations are naturally paired through the weighted $L^2$ product $(u,~v)_w=\int_{\Omega}uv \, dV/R^2.$
Thus, the factor $1/R^2$ is not imposed artificially, but naturally arises from the axisymmetric expression of the poloidal magnetic energy.

We have also checked that the adjoint formulation of Sanpei et al. \citep{Sanpei2021,Sanpei2023}, in which the geometric weight $1/R^2$ is not retained explicitly in the inner product, can be related to the present weighted formulation through a redefinition of the scalar adjoint variable. In particular, the corresponding unweighted adjoint variable is obtained by absorbing the metric factor, $\eta_{\rm unweighted} = \eta_{\rm weighted}/R^2$. Therefore, the difference between the two formulations is mainly representational rather than a limitation of the unweighted convention. The advantage of the present convention is that the self-adjoint structure
of the perturbed Grad--Shafranov operator is displayed directly.

\section{Discussion}\label{sec:Discussion}

The present reduction also provides a useful perspective on weakly non-axisymmetric equilibria. Rather than starting from the Grad--Shafranov adjoint equation and subsequently adding non-axisymmetric corrections, one may regard the three-dimensional adjoint equation as a parent formulation for organizing the axisymmetric response and the effects of weak three-dimensionality. Let $f(y)=0$, with $y=(p,\boldsymbol{J},\boldsymbol{B})$, denote the static MHD equilibrium system. The corresponding adjoint equation may be written schematically as
\begin{align*}
    Df(y)^*\Lambda = q ,
\end{align*}
where $\Lambda$ denotes the adjoint variables and $q$ is the source term determined by the derivative of the quantity of interest with respect to the equilibrium variables.

For a weakly non-axisymmetric equilibrium that admits a small-parameter expansion about an axisymmetric equilibrium \citep{plunk2020perturbing}, we formally write
\begin{align*}
    y = y_0 + \epsilon y_1 + O(\epsilon^2), \
    \Lambda = \Lambda_0 + \epsilon \Lambda_1 + O(\epsilon^2),
\end{align*}
where $0<\epsilon\ll1$, $y_0$ is an axisymmetric Grad--Shafranov equilibrium, and $y_1$ represents a three-dimensional correction to the equilibrium. At leading order, the adjoint equation is evaluated about $y_0$. In the homogeneous case considered above, restricting the leading-order adjoint variables to be axisymmetric yields a closed axisymmetric problem. The reduction derived above identifies this problem with the homogeneous Grad--Shafranov adjoint problem, together with the weighted pairing and the compatible boundary conditions. For a nonzero source term, the same interpretation applies formally when its leading-order contribution is axisymmetric.

At higher order, weak three-dimensionality may enter through the variation of the adjoint operator about $y_0$. Three-dimensional components of the source term can also contribute to the adjoint response. The three-dimensional adjoint formulation thus provides a natural framework for relating the leading-order Grad–Shafranov response to sensitivity corrections associated with weak departures from axisymmetry. This viewpoint is relevant when the leading-order equilibrium is well approximated by an axisymmetric Grad–Shafranov state, while the equilibrium, the source term, or the sensitivity of interest contains weak three-dimensional components.

\section{Conclusion}\label{sec:Conclusion}

We have examined the adjoint structure of ideal MHD equilibrium from an operator-theoretic perspective. 
Starting from the full three-dimensional equilibrium equations, we derived the corresponding adjoint system under fixed-boundary conditions and analysed its axisymmetric reduction.
The three-dimensional adjoint system is formulated without assuming magnetic flux surfaces or symmetry, and therefore provides a common formulation for both axisymmetric and fully three-dimensional equilibrium variations.
In contrast to previous work \citep{Sanpei2021,Sanpei2023}, the present approach ensures structural compatibility between axisymmetric and fully three-dimensional formulations.
This reduction shows that the resulting scalar equation coincides with the perturbed Grad–Shafranov equation. 
Equivalently, the linearised Grad–Shafranov operator arises as the axisymmetric restriction of the adjoint of the full three-dimensional equilibrium operator. 
This provides an explicit consistency statement between the three-dimensional adjoint formulation and its axisymmetric reduction.

The reduction also clarifies the correspondence between the three-dimensional adjoint variables and the reduced Grad--Shafranov variable. The scalar variable appearing in the reduced equation is identified as $\eta = -\boldsymbol{\alpha}\cdot\nabla\psi$, where $\boldsymbol{\alpha}$ is the adjoint variable associated with the force-balance equation. In the homogeneous adjoint problem considered here, $\boldsymbol{\alpha}$ is incompressible.
The reduced scalar equation is obtained directly by projecting the three-dimensional adjoint field onto the equilibrium flux gradient.

The reduced scalar problem has the same weighted operator structure as the static axisymmetric \((n=0)\) ideal-MHD perturbation problem. With the \(1/R^2\) weighted inner product, naturally associated with the axisymmetric magnetic-energy variation, the linearised Grad--Shafranov operator takes its formal self-adjoint form, although the adjoint field \(\boldsymbol{\alpha}\) is not the same object as the ideal-MHD displacement \(\boldsymbol{\xi}\).

These results suggest that the three-dimensional adjoint formulation may provide a useful framework for perturbative and sensitivity analyses around ordered non-axisymmetric equilibria. In particular, for weakly non-axisymmetric or quasi-axisymmetric configurations, the present operator-level reduction clarifies how the axisymmetric Grad--Shafranov component is embedded in the full three-dimensional adjoint problem. This viewpoint may be useful for sensitivity analysis, adjoint-based reconstruction, and bifurcation analysis of MHD equilibria \citep{Ham_2024,solano2004criticality}, where the kernel of the linearised operator and its adjoint may help characterize degeneracies, compatibility conditions, and singular responses of the linearised equilibrium system.

\begin{acknowledgements}
This work was partially supported by the ZE Research Program, Institute of Advanced Energy, Kyoto University (Reference No. ZE2025B-29). 
\end{acknowledgements}


\bibliographystyle{jpp}

\bibliography{references}

\end{document}